\documentclass[submission]{eptcs}
\providecommand{\event}{IWIGP 2012}

\usepackage{breakurl}
\usepackage{ifthen}
\usepackage{graphicx}
\usepackage{tikz}
\usepackage{tabularx}
\usetikzlibrary{arrows}
\usetikzlibrary{shapes.arrows,decorations.pathmorphing}
\usepackage{amsmath}
\usepackage[ruled, vlined]{algorithm2e}
\usepackage{xparse}
\usepackage{microtype}

\NewDocumentCommand\A{g}{\IfNoValueTF{#1}{\textbf{A}}{\textbf{A}~#1}}
\NewDocumentCommand\E{g}{\IfNoValueTF{#1}{\textbf{E}}{\textbf{E}~#1}}

\NewDocumentCommand\G{g}{\IfNoValueTF{#1}{\textbf{G}}{\textbf{G}~#1}}
\NewDocumentCommand\F{g}{\IfNoValueTF{#1}{\textbf{F}}{\textbf{F}~#1}}
\NewDocumentCommand\X{g}{\IfNoValueTF{#1}{\textbf{X}}{\textbf{X}~#1}}
\NewDocumentCommand\U{gg}{\IfNoValueTF{#1}{\textbf{U}}{#1~\textbf{U}~#2}}
\NewDocumentCommand\W{gg}{\IfNoValueTF{#1}{\textbf{W}}{#1~\textbf{W}~#2}}

\NewDocumentCommand\EG{g}{\IfNoValueTF{#1}{\textbf{EG}}{\textbf{EG}~#1}}
\NewDocumentCommand\EF{g}{\IfNoValueTF{#1}{\textbf{EF}}{\textbf{EF}~#1}}
\NewDocumentCommand\EX{g}{\IfNoValueTF{#1}{\textbf{EX}}{\textbf{EX}~#1}}
\NewDocumentCommand\EU{gg}{\IfNoValueTF{#1}{\textbf{EU}}{\textbf{E[}#1~\textbf{U}~#2\textbf{]}}}
\NewDocumentCommand\EW{gg}{\IfNoValueTF{#1}{\textbf{EW}}{\textbf{E[}#1~\textbf{W}~#2\textbf{]}}}

\NewDocumentCommand\AG{g}{\IfNoValueTF{#1}{\textbf{AG}}{\textbf{AG}~#1}}
\NewDocumentCommand\AF{g}{\IfNoValueTF{#1}{\textbf{AF}}{\textbf{AF}~#1}}
\NewDocumentCommand\AX{g}{\IfNoValueTF{#1}{\textbf{AX}}{\textbf{AX}~#1}}
\NewDocumentCommand\AU{gg}{\IfNoValueTF{#1}{\textbf{AU}}{\textbf{A[}#1~\textbf{U}~#2\textbf{]}}}
\NewDocumentCommand\AW{gg}{\IfNoValueTF{#1}{\textbf{AW}}{\textbf{A[}#1~\textbf{W}~#2\textbf{]}}}

\NewDocumentCommand\Aa{gg}{\IfNoValueTF{#1}{\ensuremath\textbf{A}_{\alpha}}{\IfNoValueTF{#2}{\ensuremath\textbf{A}_{#1}}{\ensuremath\textbf{A}_{#1}~#2}}}
\NewDocumentCommand\Ea{gg}{\IfNoValueTF{#1}{\ensuremath\textbf{E}_{\alpha}}{\IfNoValueTF{#2}{\ensuremath\textbf{E}_{#1}}{\ensuremath\textbf{E}_{#1}~#2}}}

\NewDocumentCommand\EaG{gg}{\IfNoValueTF{#1}{\ensuremath\textbf{E}_{\alpha}\textbf{G}}{\IfNoValueTF{#2}{\ensuremath\textbf{E}_{#1}\textbf{G}}{\ensuremath\textbf{E}_{#1}\textbf{G}~#2}}}
\NewDocumentCommand\EaF{gg}{\IfNoValueTF{#1}{\ensuremath\textbf{E}_{\alpha}\textbf{F}}{\IfNoValueTF{#2}{\ensuremath\textbf{E}_{#1}\textbf{F}}{\ensuremath\textbf{E}_{#1}\textbf{F}~#2}}}
\NewDocumentCommand\EaX{gg}{\IfNoValueTF{#1}{\ensuremath\textbf{E}_{\alpha}\textbf{X}}{\IfNoValueTF{#2}{\ensuremath\textbf{E}_{#1}\textbf{X}}{\ensuremath\textbf{E}_{#1}\textbf{X}~#2}}}
\NewDocumentCommand\EaU{ggg}{\IfNoValueTF{#1}{\ensuremath\textbf{E}_{\alpha}\textbf{U}}{\IfNoValueTF{#2}{\ensuremath\textbf{E}_{#1}\textbf{U}}{\ensuremath\textbf{E}_{#1}\textbf{[}#2~\textbf{U}~#3\textbf{]}}}}
\NewDocumentCommand\EaW{ggg}{\IfNoValueTF{#1}{\ensuremath\textbf{E}_{\alpha}\textbf{W}}{\IfNoValueTF{#2}{\ensuremath\textbf{E}_{#1}\textbf{W}}{\ensuremath\textbf{E}_{#1}\textbf{[}#2~\textbf{W}~#3\textbf{]}}}}

\NewDocumentCommand\AaG{gg}{\IfNoValueTF{#1}{\ensuremath\textbf{A}_{\alpha}\textbf{G}}{\IfNoValueTF{#2}{\ensuremath\textbf{A}_{#1}\textbf{G}}{\ensuremath\textbf{A}_{#1}\textbf{G}~#2}}}
\NewDocumentCommand\AaF{gg}{\IfNoValueTF{#1}{\ensuremath\textbf{A}_{\alpha}\textbf{F}}{\IfNoValueTF{#2}{\ensuremath\textbf{A}_{#1}\textbf{F}}{\ensuremath\textbf{A}_{#1}\textbf{F}~#2}}}
\NewDocumentCommand\AaX{gg}{\IfNoValueTF{#1}{\ensuremath\textbf{A}_{\alpha}\textbf{X}}{\IfNoValueTF{#2}{\ensuremath\textbf{A}_{#1}\textbf{X}}{\ensuremath\textbf{A}_{#1}\textbf{X}~#2}}}
\NewDocumentCommand\AaU{ggg}{\IfNoValueTF{#1}{\ensuremath\textbf{A}_{\alpha}\textbf{U}}{\IfNoValueTF{#2}{\ensuremath\textbf{A}_{#1}\textbf{U}}{\ensuremath\textbf{A}_{#1}\textbf{[}#2~\textbf{U}~#3\textbf{]}}}}
\NewDocumentCommand\AaW{ggg}{\IfNoValueTF{#1}{\ensuremath\textbf{A}_{\alpha}\textbf{W}}{\IfNoValueTF{#2}{\ensuremath\textbf{A}_{#1}\textbf{W}}{\ensuremath\textbf{A}_{#1}\textbf{[}#2~\textbf{W}~#3\textbf{]}}}}

\NewDocumentCommand\Kk{gg}{\IfNoValueTF{#1}{\ensuremath\textbf{K}_{ag}}{\IfNoValueTF{#2}{\ensuremath\textbf{K}_{#1}}{\ensuremath\textbf{K}_{#1}~#2}}}
\NewDocumentCommand\Ek{gg}{\IfNoValueTF{#1}{\ensuremath\textbf{E}_{g}}{\IfNoValueTF{#2}{\ensuremath\textbf{E}_{#1}}{\ensuremath\textbf{E}_{#1}~#2}}}
\NewDocumentCommand\Dk{gg}{\IfNoValueTF{#1}{\ensuremath\textbf{D}_{g}}{\IfNoValueTF{#2}{\ensuremath\textbf{D}_{#1}}{\ensuremath\textbf{D}_{#1}~#2}}}
\NewDocumentCommand\Ck{gg}{\IfNoValueTF{#1}{\ensuremath\textbf{C}_{g}}{\IfNoValueTF{#2}{\ensuremath\textbf{C}_{#1}}{\ensuremath\textbf{C}_{#1}~#2}}}

\NewDocumentCommand\Kn{gg}{\IfNoValueTF{#1}{\ensuremath\overline{\textbf{K}}_{ag}}{\IfNoValueTF{#2}{\ensuremath\overline{\textbf{K}}_{#1}}{\ensuremath\overline{\textbf{K}}_{#1}~#2}}}
\NewDocumentCommand\En{gg}{\IfNoValueTF{#1}{\ensuremath\overline{\textbf{E}}_{g}}{\IfNoValueTF{#2}{\ensuremath\overline{\textbf{E}}_{#1}}{\ensuremath\overline{\textbf{E}}_{#1}~#2}}}
\NewDocumentCommand\Dn{gg}{\IfNoValueTF{#1}{\ensuremath\overline{\textbf{D}}_{g}}{\IfNoValueTF{#2}{\ensuremath\overline{\textbf{D}}_{#1}}{\ensuremath\overline{\textbf{D}}_{#1}~#2}}}
\NewDocumentCommand\Cn{gg}{\IfNoValueTF{#1}{\ensuremath\overline{\textbf{C}}_{g}}{\IfNoValueTF{#2}{\ensuremath\overline{\textbf{C}}_{#1}}{\ensuremath\overline{\textbf{C}}_{#1}~#2}}}

\NewDocumentCommand\reachable{}{\textbf{Reachable}}

\NewDocumentCommand\Kp{gg}{\IfNoValueTF{#1}{\ensuremath\textbf{K}^{\pm}_{ag}}{\IfNoValueTF{#2}{\ensuremath\textbf{K}^{\pm}_{#1}}{\ensuremath\textbf{K}^{\pm}_{#1}~#2}}}

\newcommand{\expl}{\ensuremath{\ explains\ }}
\newcommand{\matches}{\ensuremath{\ matches\ }}

\newcommand{\toolname}{{TLACE Visualizer}}

\providecommand{\DontPrintSemicolon}{\dontprintsemicolon}

\title{Rich Counter-Examples\\for Temporal-Epistemic Logic Model Checking}
\author{
	Simon Busard\thanks{This work is supported by the European Fund for Regional Development, by the Walloon Region and by project MoVES under the Interuniversity Attraction Poles Programme --- Belgian State --- Belgian Science Policy.}
	\institute{
		ICTEAM Institute,\\
		Universit\'e catholique de Louvain,\\
		Louvain-la-Neuve, Belgium
	}
	\email{simon.busard@uclouvain.be}
	\and
	Charles Pecheur
	\institute{
		ICTEAM Institute,\\
		Universit\'e catholique de Louvain,\\
		Louvain-la-Neuve, Belgium
	}
	\email{charles.pecheur@uclouvain.be}
}
\def\titlerunning{Rich Counter-Examples for Temporal-Epistemic Logic Model Checking}
\def\authorrunning{S. Busard and C. Pecheur}

\begin{document}

	\maketitle

	\begin{abstract}
		Model checking verifies that a model of a system satisfies a given property, and otherwise produces a counter-example explaining the violation.
		The verified properties are formally expressed in temporal logics.  Some temporal logics, such as CTL, are branching: they allow to express facts about the whole computation tree of the model, rather than on each single linear computation. This branching aspect is even more critical when dealing with multi-modal logics, i.e. logics expressing facts about systems with several transition relations. A prominent example is CTLK, a logic that reasons about temporal and epistemic properties of multi-agent systems.
		In general, model checkers produce linear counter-examples for failed properties, composed of a single computation path of the model. But some branching properties are only poorly and partially explained by a linear counter-example.
		
		This paper proposes richer counter-example structures called tree-like annotated counter-examples (TLACEs), for properties in Action-Restricted CTL (ARCTL), an extension of CTL quantifying paths restricted in terms of actions labeling transitions of the model. These counter-examples have a branching structure that supports more complete description of property violations.  Elements of these counter-examples are annotated with parts of the property to give a better understanding of their structure.  Visualization and browsing of these richer counter-examples become a critical issue, as the number of branches and states can grow exponentially for deeply-nested properties.
		
		This paper formally defines the structure of TLACEs, characterizes adequate counter-examples w.r.t. models and failed properties, and gives a generation algorithm for ARCTL properties. It also illustrates the approach with examples in CTLK, using a reduction of CTLK to ARCTL.
		The proposed approach has been implemented, first by extending the NuSMV model checker to generate and export branching counter-examples, secondly by providing an interactive graphical interface to visualize and browse them.		
	\end{abstract}

	\section{Introduction}

		Model checking is a verification technique that performs an exhaustive search among states of a given finite-state machine to verify that this model satisfies a given property, expressed in temporal or richer modal logics~\cite{Clarke-Grumberg-others-99}.  
		Some of these logics, such as CTL~\cite{Clarke-Emerson-82}, are branching: they express facts about the computation tree of the model. This branching aspect is even more critical when dealing with multi-modal logics. The most common example is CTLK, a temporal-epistemic logic reasoning about time and knowledge in multi-agent systems~\cite{Penczek-Lomuscio-03}. Both temporal and epistemic information are captured as different relations over states of the model and properties express facts about all these relations.

		A major benefit of model checking is the capability to generate a counter-example when a property is not satisfied.  Unfortunately, most of the current state-of-the-art model checkers only return linear counter-examples while, in general, branching logics need branching counter-examples~\cite{Buccafurri-Eiter-others-01}.
		
		Let's take the example of Alice and Bob, where Alice randomly picks a number $N$ between $10$ and $100$ and Bob has to guess whether the number is prime or not. At each step, Bob can ask Alice whether $N$ is divisible by another number $m$. Based on Alice's answers, Bob has to say whether $N$ is prime or not.
		
		This problem can be modeled into a multi-agent system with Alice and Bob as agents; in such a model, $N$ would be undisclosed to Bob. Such a model can then be model checked to verify that Bob always finally knows whether $N$ is prime or not; this property can be expressed in CTLK as $\AF{(\Kk{Bob}{P_N} \vee \Kk{Bob}{\neg P_N})}$, where $P_N$ is true in a state in which $N$ is prime. We say that \emph{Bob knows p}, written $\Kk{Bob}{p}$, in a state $s$ if $p$ is true in all states that are undistinguishable from $s$ for Bob.
		
		This property is not verified by a model that allows Bob to ask only three questions, and a model checker checking this property would return a counter-example.
		An adequate counter-example for this property is not a single computation path: it has to show a path composed of states in which Bob does not know whether $N$ is prime, i.e. it also has to show, for each state of this path, another reachable undistinguishable state where $N$ is prime and another one where $N$ is not.
		
\begin{figure}[!ht]
			\centering
			\scalebox{0.85}{\includegraphics{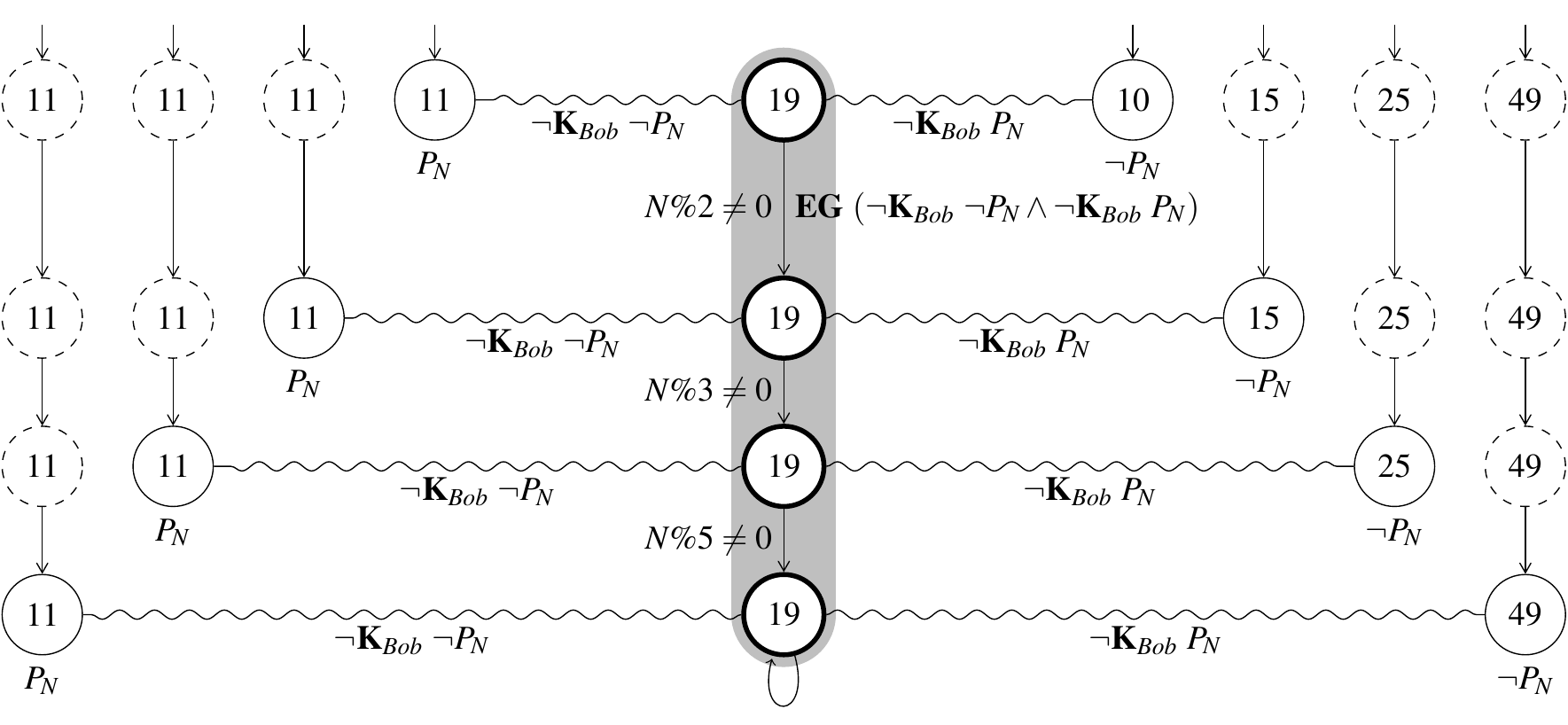}}
			\caption{A tree-like annotated counter-example for $\AF{(\Kk{Bob}{P_N} \vee \Kk{Bob}{\neg P_N})}$.}
			\label{figure:scheduler-tlace}
		\end{figure}

		Figure~\ref{figure:scheduler-tlace} gives a tree-like annotated counter-example explaining why the given property is violated. It corresponds to a scenario where $N=19$, Bob asks whether $N$ is divisible by the three first prime numbers ($2$, $3$ and $5$) and Alice answers negatively each time. The state labels are the values of $N$, which Bob does not know. The memory of Bob, used to remember Alice's answers, is not shown. The wavy transitions link together states that are undistinguishable by Bob and arrowed transitions are temporal ones. States are annotated with the properties they satisfy and transitions are annotated with properties they explain.
		
		The main branch, composed of bold states, explains how Bob asks his three questions and does not know whether $N$ is prime or not. For each state of this main path, two states that are undistinguishable by Bob from the main state are given to show that Bob does not know that $N$ is prime (right state) nor that it is not (left state). Furthermore, dashed states show that each of these states are reachable from an initial one. The highlighted path corresponds to the linear counter-example that a standard model checker would provide.
		
		This paper proposes branching structures, called \emph{tree-like annotated counter-examples} (TLACEs), that are suitable for explaining violations of branching logic properties.  Furthermore, each state of these counter-examples is annotated with the part of the (negated) property that it satisfies. These counter-examples are defined in the framework of Action-Restricted CTL (ARCTL), a branching-time logic with action-labelled transitions~\cite{Pecheur-Raimondi-07}, and their utility is illustrated in the framework of CTLK, a temporal-epistemic logic that can be reduced to ARCTL~\cite{Lomuscio-Pecheur-others-07}.
			
		Tree-like annotated counter-examples are built upon tree-like counter-examples as defined by Clarke et al.~\cite{Clarke-Jha-others-02}, which provide full counter-examples for the universal fragment of $\omega$-regular logics. These counter-examples combine cycles and finite paths in a specific, tree-like structure.
		Our tree-like annotated counter-examples extend the notion of tree-like counter-examples to ARCTL and annotate the states with formulas they satisfy to give a better understanding of the violation.
		
		TLACEs take inspiration from the work of Rasse~\cite{Rasse-92}. Rasse presents branching counter-examples for CTL interpreted over states of LTSs. Furthermore, the states of the counter-examples are annotated with the sub-formulas they explain. This notion of counter-examples is close to tree-like annotated counter-examples. Nevertheless, Rasse does not provides a way to generate counter-examples and TLACEs for ARCTL are more general since they are applicable to richer modal logics.
		
		We have extended the state-of-the-art symbolic model checker NuSMV~\cite{Cimatti-Clarke-others-02} to generate and export tree-like annotated counter-examples for ARCTL properties. 
		One of the drawbacks of these richer counter-examples is their size and complexity, which can be polynomial in the number of states of the system and exponential in the length of the checked formula in the worst case.  We have developed a tool that takes TLACEs generated by our extended NuSMV and provides a graphical interface to visualize and browse them. These tools have been used to provide a first assessment of the approach.
		They have been designed to generate and visualize TLACEs. Thanks to their parameters and functionalities, they allow the user to limit and manage the size of the counter-example.
		
		The contributions of this paper are:
		\begin{itemize}
			\item the definition of tree-like annotated counter-examples;
			\item the design of an algorithm generating these counter-examples;
			\item the implementation of this algorithm in NuSMV;
			\item the design and implementation of an interactive visualization tool for browsing these counter-examples.
		\end{itemize}
		
		This paper is structured as follows.  Section~\ref{section:branching-logics} reminds the syntax and semantics of ARCTL and CTLK.  Section~\ref{section:TLACE} defines tree-like annotated counter-examples and explains how to generate them. Section~\ref{section:example} illustrates the approach with the example of the dining cryptographers.  Section~\ref{section:tools} describes the extension of NuSMV and the visualization tool.  Section~\ref{section:evaluation} presents the evaluation of these tools.  Finally, Section~\ref{section:related-work} presents related work.

	\section{Temporal and Epistemic Logics}
	\label{section:branching-logics}
	
		This section presents the logics used in this work, ARCTL and CTLK. It first presents the syntax and semantics of ARCTL~\cite{Pecheur-Raimondi-07}, then presents CTLK, a temporal-epistemic logic, and describes a reduction of CTLK to ARCTL.

		\subsection{Action-Restricted CTL}
		
			Action-Restricted CTL is an extension of CTL applied to systems with labelled states and actions, where temporal operators are augmented with propositional expressions over actions, expressing properties of particular paths of the system. 
			
			In addition to the usual logical connectives ($\neg$, $\vee$, $\wedge$, $\implies$ and $\iff$), ARCTL provides temporal operators, composed of an action-restricted path quantifier $\Ea$ or $\Aa$ immediately followed by a path operator ($\X$, $\G$, $\F$, $\U$ and $\W$). Path operators define path formulas while path quantifiers and logical connectives define state formulas. Actions expressions $\alpha$ are composed of actions and logical connectives.
			For example, $\EaX{a}{\phi}$ means that there exists a successor reachable through the action $a$ that satisfies $\phi$; $\AaG{b}{\psi}$ means that all states reachable through $b$ actions satisfy $\psi$.
			
			ARCTL properties are interpreted over the states of a Mixed Transition System (MTS).  An MTS is a tuple $\mathcal{M} = (S, S_{0}, A, T, \mathcal{V}_S, \mathcal{V}_A)$ over two sets of atomic propositions $P_S$ and $P_A$, where $S$ is a set of states, $S_{0} \subseteq S$ are initial states, $A$ is a set of actions, $T \subseteq S \times A \times S$ is a transition relation, and $\mathcal{V}_S : S \rightarrow 2^{P_S}$ and $\mathcal{V}_A : A \rightarrow 2^{P_A}$ are two functions labeling states with subsets of $P_S$, and labeling actions with subsets of $P_A$, respectively. These two functions represent the propositions that are interpreted over states and actions, respectively.  We write $s \xrightarrow{a} s'$ for $(s, a, s') \in T$.
			
		A path of $\mathcal{M}$ starting at $s_0$ is a (finite or infinite) sequence of states and actions $w = \langle s_0, a_1, s_1, a_2, s_2,... \rangle$ such that $s_i \xrightarrow{a_{i+1}} s_{i+1}$; $w(i)$ denotes $s_i$. 
		$\Pi(\mathcal{M}, s)$ is the set of maximal paths in $\mathcal{M}$ starting at $s$.
		$\mathcal{M}|_{\alpha} = (S, S_{0}, A, T|_{\alpha}, \mathcal{V}_S, \mathcal{V}_A)$ is the $\alpha$-restriction of $\mathcal{M}$ where $T|_{\alpha} = \{(s,a,s') \in T ~|~ \mathcal{M}, a \models \alpha\}$ is the transition relation $T$ where only actions $a$ satisfying $\alpha$ are considered.
		
		We write $\mathcal{M}, s \models \phi$ when a state $s$ of an MTS $\mathcal{M}$ satisfies an ARCTL property $\phi$.  Logical connectives are interpreted in the natural way.  For temporal operators, $s$ satisfies $\Ea{\alpha}{\pi}$ (resp. $\Aa{\alpha}{\pi}$), where $\pi$ is a path formula, if and only if there exists a path (resp. all paths) in $\Pi(\mathcal{M}|_{\alpha}, s)$ satisfying $\pi$.
		
		A path $w$ of $\mathcal{M}$ satisfies $\X{\phi}$ if and only if $w(1)$ satisfies $\phi$; $w$ satisfies $\F{\phi}$ (resp. $\G{\phi}$) iff $w(i)$ satisfies $\phi$ for some (resp. for all) $i$. Finally, $w$ satisfies $\U{\phi}{\psi}$ iff $w(i)$ satisfies $\psi$ for some $i$ and $w(j)$ satisfies $\phi$ for all $j < i$. $\W{\phi}{\psi}$ is equivalent to $(\U{\phi}{\psi}) \lor (\G{\phi})$.	
		
		In the remainder of this paper, all given ARCTL formulas are considered as reduced to their \emph{negative normal form}: negations are distributed over all operators so that they are only applied to atomic propositions.  Furthermore, equivalences can be applied to reduce formulas to the following base cases: $b$, $\neg b$ (for atomic propositions $b$), $\phi \wedge \psi$, $\phi \vee \psi$, $\EaX{\alpha}{\phi}$, $\EaG{\alpha}{\phi}$, $\EaU{\alpha}{\phi}{\psi}$ and $\Aa{\alpha}{\pi}$. This allows a more concise presentation of concepts without loss of generality.

		\subsection{CTLK}
		
			CTLK is a branching-time epistemic logic mixing knowledge relations and temporal ones~\cite{Penczek-Lomuscio-03}. This logic is designed to express facts about time and knowledge of agents in a multi-agent system. This section presents the syntax and semantics of CTLK.
		
			CTLK provides the usual logical connectives together with CTL operators ($\EX$, $\AG$, $\EF$, etc.) and the knowledge operator $\Kk$ where $ag$ is an agent. It also provides some other epistemic operators for the \emph{group knowledge} $\Ek$, the \emph{distributed knowledge} $\Dk$ and the \emph{common knowledge} $\Ck$, where $g$ is a group of agents, but they are not developed here. Nevertheless, they also can be reduced to ARCTL~\cite{Lomuscio-Pecheur-others-07}.
			
			CTLK is interpreted over multi-agent systems, where each agent is aware of the possible behaviors of the system and of its own local state but not of the local state of other agents.  Formally, a multi-agent system composed of $n$ agents is a Kripke structure $\mathcal{M_A} = (S, S_0, T, \sim_1, ..., \sim_n, \mathcal{V})$ where $T \subseteq S \times S$ is a (temporal) transition relation and $\sim_i\; \subseteq S \times S$ are epistemic relations.  $(s,s') \in\; \sim_i$, written $s \sim_i s'$, iff $s$ and $s'$ are reachable states that share the same local state for agent $ag_i$.
			
			An agent $ag_i$ knows $\phi$ in a state $s$ iff $\phi$ holds in all reachable states that are undistinguishable from $s$ by $ag_i$.  Formally, $\mathcal{M_A}, s \models \Kk{ag_i}{\phi}$ if and only if $\forall s' \in S : s' \sim_i s \Rightarrow \mathcal{M_A}, s' \models \phi$.
			Note that $\sim_i$ must be restricted to \emph{reachable} states (i.e. $T^*(S_0)$), capturing the fact that $ag_i$ knows the global system behavior.  A witness for a reachable state is thus a reverse execution path back to an initial state.

		\subsection{From CTLK to ARCTL}
		
			Some multi-modal branching logics, i.e. logics dealing with more than one transition relation, can be reduced to ARCTL. CTLK is such a logic: a multi-agent system and a CTLK formula can be reduced to an MTS and an ARCTL formula, respectively~\cite{Lomuscio-Pecheur-others-07}.
			Given a multi-agent system, the corresponding MTS has the same set of states. The set of actions contains actions $RUN$ and $BACK$, used to label temporal and reverse temporal transitions, and one action per agent, to label epistemic transitions. The transition relation is an aggregation of the temporal relation, the reverse temporal relation and the epistemic ones, using corresponding actions. The labeling of states is augmented with the proposition $Init$ to label initial states. $Init$ is used to express the reachability of states: a state is reachable from an initial state iff it satisfies $\EaF{\{BACK\}}{Init}$.
			
			Formally, given a multi-agent system $\mathcal{M_A} = (S, S_0, T, \sim_1, ..., \sim_n, \mathcal{V})$ composed of $n$ agents, the corresponding MTS is given by $\mathcal{M} = (S, S_0, A, T', \mathcal{V}_S, \mathcal{V}_A)$ where
			\begin{itemize}
				\item $A = 2^{\{RUN, BACK, Agt_1, ..., Agt_n\}}$\footnote{The use of subsets of labels is needed to handle distributed knowledge. See~\cite{Lomuscio-Pecheur-others-07} for details.}
				\item for all states $s, s' \in S$ : (i) $(s, \{RUN\}, s') \in T'$ iff $(s, s') \in T$; (ii) $(s, \{BACK\}, s') \in T'$ iff $(s', s) \in T$; (iii) $(s, \{Agt_i\}, s') \in T'$ iff $s \sim_i s'$; (iv) $(s, \{Agt_i ~|~ ag_i \in g\}, s') \in T'$ iff $\forall ag_i \in g : s \sim_i s'$.
				\item $\mathcal{V}_S(s) = \mathcal{V}(s) \cup \{Init\} \textrm{ if } s \in S_0, \mathcal{V}(s) \textrm{ otherwise}$; $\mathcal{V}_A$ is the identity function.
			\end{itemize} 
			
			To reduce a CTLK formula into an ARCTL one, we use the labels $RUN$, $BACK$ and $Agt_i$ to represent a temporal transition, a reverse temporal transition and an epistemic transition of agent $ag_i$, respectively. Formally, let the function $R$ reduce CTLK formulas into their ARCTL form, $R$ is inductively defined as
			\begin{itemize}
				\item $R(b) = b$ if $b$ is a propositional formula;
				\item $R(\EX{\phi}) = \EaX{\{RUN\}}{R(\phi)}$; $R(\EG{\phi}) = \EaG{\{RUN\}}{R(\phi)}$; $R(\EU{\phi}{\psi}) = \EaU{\{RUN\}}{R(\phi)}{R(\psi)}$;
				\item $R(\Kk{a_i}{\phi}) = \AaX{\{Agt_i\}}{(\reachable \implies R(\phi))}$, where $\reachable$ is a shortcut for $\EaF{\{BACK\}}{Init}$.
			\end{itemize} 
		
	\section{Tree-Like Annotated Counter-Examples}
	\label{section:TLACE}
	
		This section presents generic structures called \emph{tree-like annotated counter-examples}, or TLACEs for short, for explaining why a state $s$ of a system does not satisfy an ARCTL property $\phi$.  A \emph{counter-example} explaining a violation of $\phi$ in $s$ amounts to a \emph{witness} explaining satisfaction of $\neg\phi$ in $s$.  From now on, this paper will discuss TLACEs as witnesses of properties rather than counter-examples to avoid carrying negations all through.
		
		A TLACE witnessing $\phi$ is a \emph{node} composed of a state annotated with the direct sub-formulas of $\phi$ that it satisfies. Furthermore, for each existential temporal sub-formula it satisfies (i.e. $\Ea$ formulas), the node contains a branch explaining the formula. A branch is a list of nodes and actions representing a path in the model and witnessing the temporal formula.
		Formally, tree-like annotated counter-examples are defined based on the following grammar of nodes $n$ and paths $p$:
		\begin{align*}
			n 	& ::= node(s, \{(b ~|~ \neg b)^{*}\}, \{(\Ea{\alpha}{\pi} : p)^{*}\}, \{(\Aa{\alpha}{\pi})^{*}\}) \\
			p	& ::= \langle n, (a, n)^* \rangle ~|~ \langle n, (a, n)^*, a, loop(n) \rangle
		\end{align*}
		where $s$ are states, $b$ are atomic propositions, $a$ are actions, $\alpha$ are boolean expressions over actions and $\pi$ are ARCTL path formulas. The $loop$ marker is used to represent a looping path, the marked node is the first one of the loop.  Given a node $node(s, aps, \{\Ea{\alpha_i}{\pi_i} : p_i\}, abs)$, each $\Ea{\alpha_i}{\pi_i} : p_i$ is called a \textit{branch}; $aps$, $\{\Ea{\alpha_i}{\pi_i}\}$ and $abs$ are \textit{annotations}. Let $State(node(s, aps, ebs, abs)) = s$ and $First(\langle n_0, a_1, ..., n_m \rangle) = First(\langle n_0, a_1, ..., n_m, a_{m+1}, loop(n') \rangle) = n_0$.\\A TLACE node $node(s, aps, \{\Ea{\alpha_i}{\pi_i} : p_i\}, abs)$ is \emph{consistent} iff all its paths $p_i$ are \emph{consistent} and satisfy $s = State(First(p_i))$; a TLACE path $\langle n_0, a_1, ..., n_m \rangle$ is \emph{consistent} iff all its nodes are \emph{consistent}; a TLACE path $\langle n_0, a_1, ..., n_m, a_{m+1}, loop(n') \rangle$ is \emph{consistent} iff $\langle n_0, a_1, ..., n_m, a_{m+1}, n' \rangle$ is \emph{consistent} and $n' = n_j$ for some $0 \leq j \leq m$.  We will only consider consistent TLACEs in the sequel, and call TLACE, or witness, a consistent TLACE node.

		\subsection{Adequate Witnesses}
		\label{section:adequate}

		TLACEs ought to be \emph{adequate} witnesses for a formula $\phi$ in a state $s$ of a model $\mathcal{M}$, in a precisely defined sense.  Given a tree-like annotated witness $n = node(s, aps, ebs, abs)$, 	the witness has to satisfy the following conditions to be adequate:
		\begin{itemize}
			\item The witness represents a part of the computation tree of $\mathcal{M}$. Its paths are execution paths in $\mathcal{M}$.
			\item The atomic propositions annotating nodes of the witness are satisfied in the corresponding states of $\mathcal{M}$ and the actions of paths satisfy action formulas of $\phi$.
			\item The witness is effectively a witness for $\phi$. It represents (generally partially) a computation tree ensuring $\phi$.
			\item The annotations of the witness are coherent with $\phi$.  Branch annotations are sub-formulas of $\phi$.
		\end{itemize}
		
		The first condition above is formally expressed as $n \matches \mathcal{M}$---the witness is part of the model---while the three last ones are expressed as $n \expl (\mathcal{M}, \phi)$---the witness explains the property in the model.
		An \emph{adequate} witness for $\phi$ in $s$ of $\mathcal{M}$ is a witness in $s$ that matches $\mathcal{M}$ and explains $\phi$ in $\mathcal{M}$.
		
	 The witness $n$ matches $\mathcal{M}$ if $s$ is a state of $\mathcal{M}$ and each path in $ebs$ corresponds to a path in $\mathcal{M}$ such that the nodes recursively match their respective states.
	 
		Formally, let $\mathcal{M} = (S, S_0, A, T, \mathcal{V}_S, \mathcal{V}_A)$.
		A node $n = (s, aps, ebs, abs) \matches \mathcal{M}$ iff (i) $s \in S$ and (ii) $\forall (\Ea{\alpha_i}{\pi_i} : p_i) \in ebs : p_i \matches \mathcal{M}$.  A path $p = \langle n_0, a_1, ..., n_m \rangle \matches \mathcal{M}$ iff (i) $\forall i, 0 \leq i \leq m : n_i \matches \mathcal{M}$ and (ii) $\forall i, 0 \leq i < m : State(n_i) \xrightarrow{a_{i+1}} State(n_{i+1})$. A looping path $p = \langle n_0, a_1, ..., $ $ n_m, a_{m+1}, loop(n') \rangle \matches \mathcal{M}$ iff $\langle n_0, a_1, ..., n_m, a_{m+1}, n' \rangle$ $\matches \mathcal{M}$.
		
		The witness $n$ explains $\phi$ in $\mathcal{M}$ if it has the shape of a witness for $\phi$.  This highly depends on the structure of $\phi$.  For example, a witness for $\phi_1 \land \phi_2$ is a node composed of the annotations and branches of two nodes explaining $\phi_1$ and $\phi_2$, respectively.
		
		Formally, $n \expl (\mathcal{M}, \phi)$ is defined recursively over the structure of $\phi$. This definition is given for state formulas $\phi$ by the following two tables.
				
		\begin{center}
		\begin{tabularx}{\textwidth}{|p{2cm}|X|}
			\hline
			$\phi$							& $n \expl (\mathcal{M}, \phi)$ iff\dots		\\
			\hline
			$true$							& $n = node(s, \{\}, \{\}, \{\})$	\\
			$b$ or $\neg b$				& $n = node(s, \{\phi\}, \{\}, \{\})$ and $\mathcal{M}, s \models \phi$ \\
			$\phi_1 \vee \phi_2$		& $n \expl (\mathcal{M}, \phi_1)$ or $n \expl (\mathcal{M}, \phi_2)$	\\
			$\phi_1 \wedge \phi_2$	& $n = node(s, aps_1 \cup aps_2, ebs_1 \cup ebs_2, abs_1 \cup abs_2)$ \\
											& and $node(s, aps_i, ebs_i, abs_i) \expl (\mathcal{M}, \phi_i)$ \\		
		\end{tabularx}\vspace{-1pt}
		\begin{tabularx}{\textwidth}{|p{2cm}|X|}
			$\Aa{\alpha}{\pi}$			& $n = node(s, \{\}, \{\}, \{\Aa{\alpha}{\pi}\})$	\\
			$\Ea{\alpha}{\pi}$			& $n = node(s, \{\}, \{\Ea{\alpha}{\pi} : p\}, \{\})$ and $p \expl (\mathcal{M}, \phi)$ \\
			\hline
		\end{tabularx}
		\begin{tabularx}{\textwidth}{|p{2cm}|X|}
			\hline
			$\phi$							& $p \expl (\mathcal{M}, \phi)$ iff\dots		\\
			\hline
			$\EaX{\alpha}{\phi}$		& $p = \langle node(s, \{\}, \{\}, \{\}), a, n_1 \rangle$,
												$n_1 \expl (\mathcal{M}, \phi)$ and $\mathcal{M}, a \models \alpha$ \\
			$\EaU{\alpha}{\phi_1}{\phi_2}$		& $p = \langle n_0, a_1, ..., n_m \rangle$,
												$\forall i, 0 \leq i < m : n_i \expl (\mathcal{M}, \phi_1)$,
												$n_m \expl (\mathcal{M}, \phi_2)$ \\
											& and $\forall i, 1 \leq i \leq m : \mathcal{M}, a_i \models \alpha$ \\
			$\EaG{\alpha}{\phi}$		& $p = \langle n_0, a_1, ..., n_m, a_{m+1}, loop(n') \rangle$,
												$\forall i : n_i \expl (\mathcal{M}, \phi)$	\\
											& and $\forall i, 1 \leq i \leq m+1 : \mathcal{M}, a_i \models \alpha$ \\
			\hline
		\end{tabularx}
		\end{center}

		Finally, $n$ is an adequate witness for $\mathcal{M}, s \models \phi$ iff $State(n) = s$, $n \matches \mathcal{M}$ and $n \expl (\mathcal{M}, \phi)$.
		
		Note that universal temporal sub-formulas (i.e. $\Aa$ formulas) are not explained: elements of $abs$ are just annotations, i.e. ARCTL formulas. While explaining an $\Ea$ formula needs only one path, the whole system is potentially needed to explain an $\Aa$ formula. There is no inherent difficulty in explaining them, but it would usually result in a huge, unmanageable structure.
		
		Tree-like annotated witnesses are full witnesses for the existential fragment of ARCTL. In this fragment, only existential path quantifiers are allowed and negations are only applicable to atomic propositions. TLACEs are full witnesses in the sense that there exists an \emph{adequate} witness for $\mathcal{M}, s \models \phi$ if and only if $\phi$ is satisfied in the state $s$ of $\mathcal{M}$. They adequately describe the satisfaction because they provide all the reasonably useful information. On the other hand, because tree-like annotated witnesses do not explain universal operators, they are not full witnesses for full ARCTL.

		\subsection{Generating Counter-Examples}
		\label{section:generating}
			
		This section gives an algorithm to generate tree-like annotated witnesses.
		This algorithm is described by the function $explain$ given below, which takes as arguments a mixed transition system $\mathcal{M} = (S, S_0, A, T, \mathcal{V}_S, \mathcal{V}_A)$, a state $s \in S$, and a property $\phi$ such that $\mathcal{M}, s \models \phi$, and returns a consistent and adequate tree-like annotated witness for $\mathcal{M}, s \models \phi$.  To perform this computation, it works recursively on the structure of $\phi$.
		
		The algorithm uses the sub-algorithms $EaGexplain$, $EaUexplain$ and $EaXexplain$, which return paths in $\mathcal{M}$ satisfying $\EaG$, $\EaU$ and $\EaX$ operators, respectively.
		More precisely, 
		$EaGexplain(\mathcal{M}, s, \phi, \alpha)$ returns a path $\langle s_0, a_1, ..., s_m \rangle$ where $s = s_0$, $\forall i, 0 \leq i \leq m : \mathcal{M}, s_i \models \phi$, $\exists k, 0 \leq k < m : s_k = s_m$ and  $\forall i, 1 \leq i \leq m : \mathcal{M}, a_i \models \alpha$.
		$EaUexplain(\mathcal{M}, s, \phi, \psi, \alpha)$ returns a path $\langle s_0, a_1, ..., s_m \rangle$ where $s = s_0$, $\forall i, 0 \leq i < m :\mathcal{M}, s_i \models \phi$, $\mathcal{M}, s_m \models \psi$ and $\forall i, 1 \leq i \leq m : \mathcal{M}, a_i \models \alpha$.
		$EaXexplain(\mathcal{M}, s, \phi, \alpha)$ returns a path $\langle s_0, a_1, s_1 \rangle$ where $s = s_0$, $\mathcal{M}, s_1 \models \phi$ and $\mathcal{M}, a_1 \models \alpha$.
		
		\begin{function}[!ht]
			\DontPrintSemicolon
			\KwData{$\mathcal{M}$ a Mixed Transition System, $s$ a state of $\mathcal{M}$, $\phi$ an ARCTL property, s.t. $\mathcal{M}, s \models \phi$.}
			\KwResult{a tree-like annotated witness $n$ s.t. $State(n) = s$, $n \matches \mathcal{M}$ and $n \expl (\mathcal{M}, \phi)$.}
			\BlankLine
			\Switch{$\phi$}{
				\Case{$true$}{
					\Return{$node(s, \{\}, \{\}, \{\})$}
				}
				\Case{$b$, $\neg b$}{
					\Return{$node(s, \{\phi\}, \{\}, \{\})$}
				}
				\Case{$\psi_1 \vee \psi_2$}{
					\lIf{$\mathcal{M}, s \models \psi_1$}{\Return{$explain(\mathcal{M}, s, \psi_1)$}}\;
					\lElse{\Return{$explain(\mathcal{M}, s, \psi_2)$}}
				}
				\Case{$\psi_1 \wedge \psi_2$}{
					$node(s, aps_1, ebs_1, abs_1) \leftarrow explain(\mathcal{M}, s, \psi_1)$\;
					$node(s, aps_2, ebs_2, abs_2) \leftarrow explain(\mathcal{M}, s, \psi_2)$\;
					\Return{$node(s, aps_1 \cup aps_2, ebs_1 \cup ebs_2, abs_1 \cup abs_2)$}
				}
				\Case{$\Aa{\alpha}{\pi}$}{
					\Return{$node(s, \{\}, \{\}, \{\Aa{\alpha}{\pi}\})$}
				}
				\Case{$\EaX{\alpha}{\psi}$}{
					$\langle s_0, a_1, s_1 \rangle \leftarrow EaXexplain(\mathcal{M}, s, \psi, \alpha)$\;
					\Return{$node(s, \{\}, \{\EaX{\alpha}{\psi} : \langle node(s_0, \{\}, \{\}, \{\}), a_1, explain(\mathcal{M}, s_1, \psi) \rangle\}, \{\})$}
				}
				\Case{$\EaU{\alpha}{\psi_1}{\psi_2}$}{
					$\langle s_0, a_1, ..., s_m \rangle \leftarrow EaUexplain(\mathcal{M}, s, \psi_1, \psi_2, \alpha)$\;
					$p \leftarrow \langle \rangle$\;
					\For{$i \in 0 .. m-1$}{
						$p \leftarrow p + \langle explain(\mathcal{M}, s_i, \psi_1), a_{i+1}\rangle$
					}
					\Return{$node(s, \{\}, \{\EaU{\alpha}{\psi_1}{\psi_2} : p + \langle explain(\mathcal{M}, s_m, \psi_2)\rangle\}, \{\})$}
				}
				\Case{$\EaG{\alpha}{\psi}$}{
					$\langle s_0, a_1, ..., s_m \rangle \leftarrow EGexplain(\mathcal{M}, s, \psi, \alpha)$\;
					$p \leftarrow \langle \rangle$\;
					\For{$i \in 0 .. m-1$}{
						$n_i \leftarrow explain(\mathcal{M}, s_i, \psi)$\;
						\lIf{$s_i = s_m$}{$n' \leftarrow n_i$}\;
						$p \leftarrow p + \langle n_i, a_{i+1} \rangle$
					}
					\Return{$node(s, \{\}, \{\EaG{\alpha}{\psi} : p + \langle loop(n')\rangle\}, \{\})$}
				}
			}
			\caption{explain($\mathcal{M}$, $s$, $\phi$)}
			\label{function:explain}
		\end{function}
			
		This algorithm is correct, i.e. if $\mathcal{M}, s \models \phi$, then $explain(\mathcal{M}, s, \phi)$ returns a \emph{consistent} and \emph{adequate} tree-like annotated witness for $\mathcal{M}, s \models \phi$. Its correctness can be proved using induction over the structure of $\phi$. Due to space limit, the proof is not developed here, but the intuition is given for the $\EaU$ case.
		First, $EaUexplain$ returns a witness path for $\EaU{\alpha}{\psi_1}{\psi_2}$, so $\mathcal{M}, s_m \models \psi_2$, $\mathcal{M}, s_i \models \psi_1$ for $i < m$ and $\mathcal{M}, a_i \models \alpha$ for all $i$. 
		The construction of $p$ in the \emph{for} loop and the following instruction
		builds a path composed of nodes witnessing $\psi_1$ with a last node witnessing $\psi_2$, thus altogether $p$ correctly explains $\EaU{\alpha}{\psi_1}{\psi_2}$ and the result explains $\EaU{\alpha}{\psi_1}{\psi_2}$. The result clearly belongs to $\mathcal{M}$ since $EaUexplain$ returns a path in $\mathcal{M}$. Finally, by construction, the state of the result is $s$.
		
	\newpage	
	\section{Example: the Dining Cryptographers Protocol}
	\label{section:example}
	
		This section uses the dining cryptographers protocol~\cite{Chaum-88}, a well-known example in temporal epistemic logic, to illustrate the applicability of tree-like annotated counter-examples to CTLK and multi-agent systems.  A model of the protocol is given, a CTLK property violated by the system is presented and the corresponding counter-example is described.
		
		Summarizing~\cite{Chaum-88}, the protocol of the dining cryptographers can be described as follows:
		\begin{quote}
			Three cryptographers at restaurant made an arrangement with the waiter for the bill to be paid anonymously. One of them might be the payer or it might be NSA. The three cryptographers wonder if NSA is paying but nobody wants to say if she pays.  To resolve this problem, the following protocol is performed. Each cryptographer flips a coin behind his menu such that her right neighbor and she can see the result. Each one then claims aloud whether the two coins that she saw are equal or different, stating the opposite if she is the payer. An odd number of differences indicates a lier, and then a payer; an even number indicates that NSA is paying, assuming that the dinner was paid once. No not-paying cryptographer can tell which one of the others is the payer.
		\end{quote}
		
		We consider a model composed of three agents $a$, $b$ and $c$, representing three cryptographers.  Each agent knows whether she paid, the result of the coin flips to her left and right and the claims of all agents.
		The protocol is executed in three steps. The initial step determines, for each agent, if she is the payer or not, making sure that at most one of them is the payer. Then each agent flips her coin.  Finally each agent makes her claim, depending on the results of the coin flips and whether she is the payer or not.
		
		We consider the CTLK property $\phi \equiv \neg a.payer \implies \AF{(\Kk{a}{b.payer} \vee \Kk{a}{c.payer})}$.  It expresses that if $a$ is not the payer, then she will eventually either know that $b$ is the payer or that $c$ is. This property is obviously violated by the system as the protocol ensures anonymity of the payer.
		
\begin{figure}[!ht]
			\center
			\scalebox{0.83}{\includegraphics{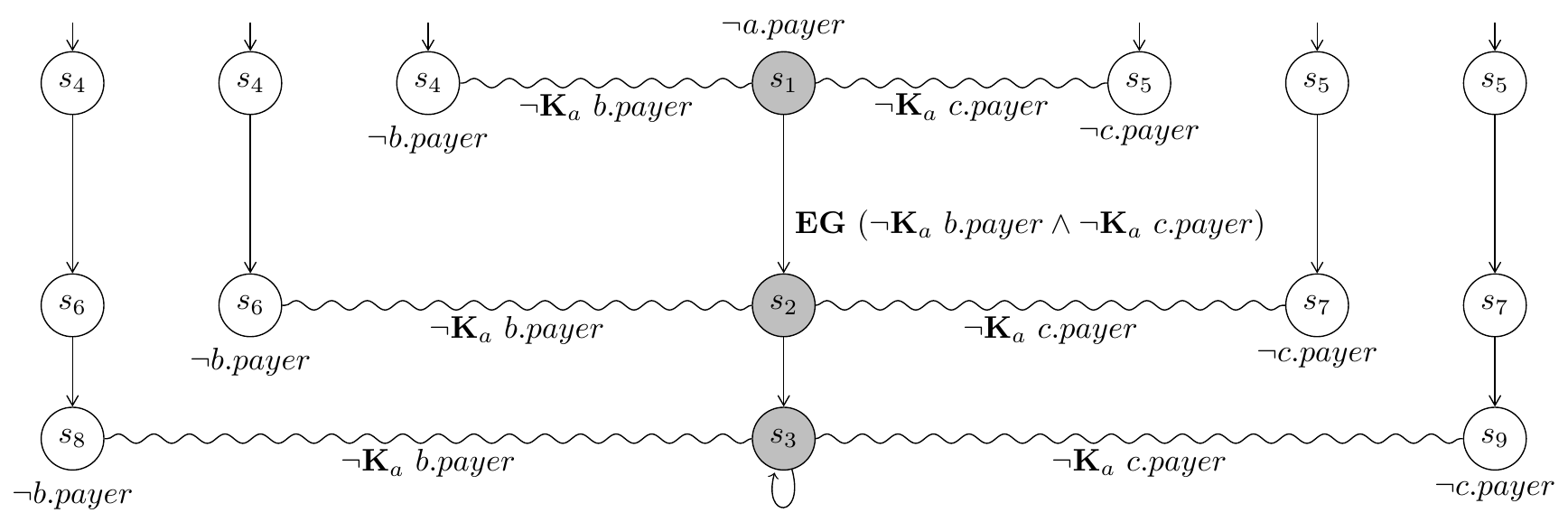}}
			\caption{A counter-example for 
			$\neg a.payer \implies \protect\AF{(\protect\Kk{a}{b.payer} \vee \protect\Kk{a}{c.payer})}$ 
			in the model of the dining cryptographers.
			Straight arrows are temporal transitions, waived arrows are epistemic equivalences for $a$.
			}
			\label{figure:crypto-tlace}
		\end{figure}

		The violation of this property is explained by the tree-like annotated counter-example presented in Figure \ref{figure:crypto-tlace}.
		It is a witness for $\neg \phi \equiv \neg a.payer \wedge \EG{(\neg\Kk{a}{b.payer} \wedge \neg\Kk{a}{c.payer})}$, with  a path ending with a loop (the gray states) composed of states satisfying $\neg\Kk{a}{b.payer} \wedge \neg\Kk{a}{c.payer}$. Each state explains $\neg\Kk{a}{b.payer}$ by giving an equivalent state for $a$ satisfying $\neg b.payer$ and with a backward path to an initial state (and similarly for $\neg\Kk{a}{c.payer}$).

		This counter-example clearly illustrates the need for rich counter-examples. A linear counter-example would only give the gray part of the presented counter-example, without the annotations, missing a lot of information, and the understanding of the violation would be very hard for the user. This counter-example is a good representative of CTLK properties mixing temporal and epistemic operators.

	\section{Implementation}
	\label{section:tools}

		The principles exposed in the previous sections have been implemented in two distinct parts. First, the well-known open-source symbolic model checker NuSMV~\cite{Cimatti-Clarke-others-02} has been modified to generate tree-like annotated counter-examples for ARCTL properties. Second, a new tool called \toolname{} has been implemented to visualize and manipulate these counter-examples\footnote{These tools are available on http://lvl.info.ucl.ac.be/Tools/NuSMV-ARCTL-TLACE.}. An XML format has been designed as a transfer syntax between the two tools.
		
		The modified version of NuSMV	implements the generation algorithm presented in Section \ref{section:generating}. The implementation is based on the $EXexplain$, $EUexplain$ and $EGexplain$ algorithms already implemented in NuSMV and modified to take actions into account. It generates tree-like annotated counter-examples and can export them into a custom XML format.
		
		Technically, the algorithm has been implemented slightly differently than presented in this paper but the result is equivalent. The implemented algorithm generates a witness for every ARCTL formula, without prior reduction to normal form. Negations are handled within the recursive traversal of sub-formulas.
		
		The implementation supports two kinds of parameters to limit the amount of generated information.  The first set of parameters allows to selectively generate branches only for some temporal operators (e.g. for $\EaX$ but not for $\EaU$ nor $\EaG$). The second parameter limits the maximum depth of generated branches, in terms of number of nested temporal operators.
		
\begin{figure}[!ht]
			\centering
			\includegraphics[width=\textwidth]{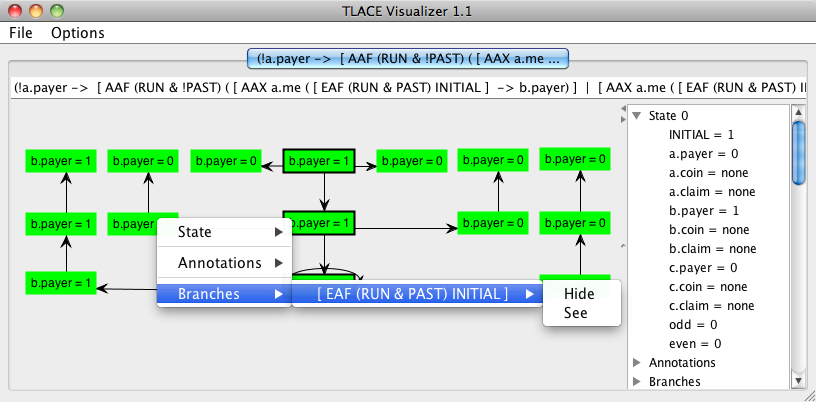}
			\caption{A snapshot of the interface illustrating the different feature of the tool.}
			\label{figure:snapshot}
		\end{figure}

		In terms of output, the original version of NuSMV returns only linear (looping) paths as counter-examples; on the other hand, the modified version of NuSMV returns richer information that becomes difficult to display in a text format.
		We developed \toolname{}, an interactive graphical interface application for displaying and browsing tree-like annotated counter-examples.  The counter-examples are loaded from XML files produced by the modified NuSMV and pictured as a graph in the main area of the interface.  The tool also provides different means to arrange the layout of the graph and explore the detailed information associated to each node.  A snapshot of the interface is given in Figure~\ref{figure:snapshot}.
		
		The tool automatically lays out the counter-example upon loading, according to a custom layout algorithm that takes into account the semantic structure of the counter-example.  This representation presents the general structure of the counter-example, showing branches and loops. 
			Single states or entire subtrees can be dragged around for better readability.
			To support browsing of larger graphs, branches can be folded and unfolded to reduce clutter and selectively show relevant information.
			
		A side panel displays the values of all variables and annotations along a selected path in the graph, in a collapsible hierarchical presentation. All variable values can also be accessed as a pop-up menu on each node in the main panel, and variables can be selected for display as part of the node's label, giving immediate visibility for a few variables of interest.
		
		Both our extended NuSMV and the visualization tool currently only support ARCTL logic natively.  That means that epistemic relations are shown in their ARCTL-reduced form.  A desirable future extension is to display counter-examples according to their original logic notations.  This requires some additional engineering but poses no major technical challenge.
	
	\section{Evaluation}
	\label{section:evaluation}
	
		This section first assesses the benefits of the provided browsing facilities to manage complex counter-examples in the context of multi-agent systems and CTLK.  It then discusses how the approach could be extended to handle larger counter-examples and universal witnesses using an interactive, incremental generation.
		
		\subsection{Richness and Complexity of TLACEs} 
		
		The need for branching counter-examples for CTLK properties has already been illustrated in Section~\ref{section:example} with a property violated by the protocol of the dining cryptographers. This example showed that a linear counter-example was not enough to fully understand why the model (or the property) was wrong.
		
		To illustrate the increasing richness and complexity of counter-examples, let's consider the following property on the dining cryptographers: cryptographer $a$ will eventually know whether cryptographer $b$ knows whether $a$ is paying or not. Let $\Kp{ag}{\phi}$ be a shortcut for $\Kk{ag}{\phi} \vee \Kk{ag}{\neg \phi}$, meaning that agent $ag$ knows whether $\phi$ is true or not. The above property can then be expressed as $\AF{\Kp{a}{\Kp{b}{a.payer}}}$, meaning that $a$ always eventually knows whether $b$ knows if $a$ is the payer or not. This property is violated by the model since, if $b$ or $c$ is the payer, then $a$ cannot say whether $b$ knows whether $a$ paid or not (if $b$ paid, $b$ knows, if $c$ paid, $b$ does not).
		
\begin{figure}[!ht]
			\centering
			\includegraphics[width=\textwidth]{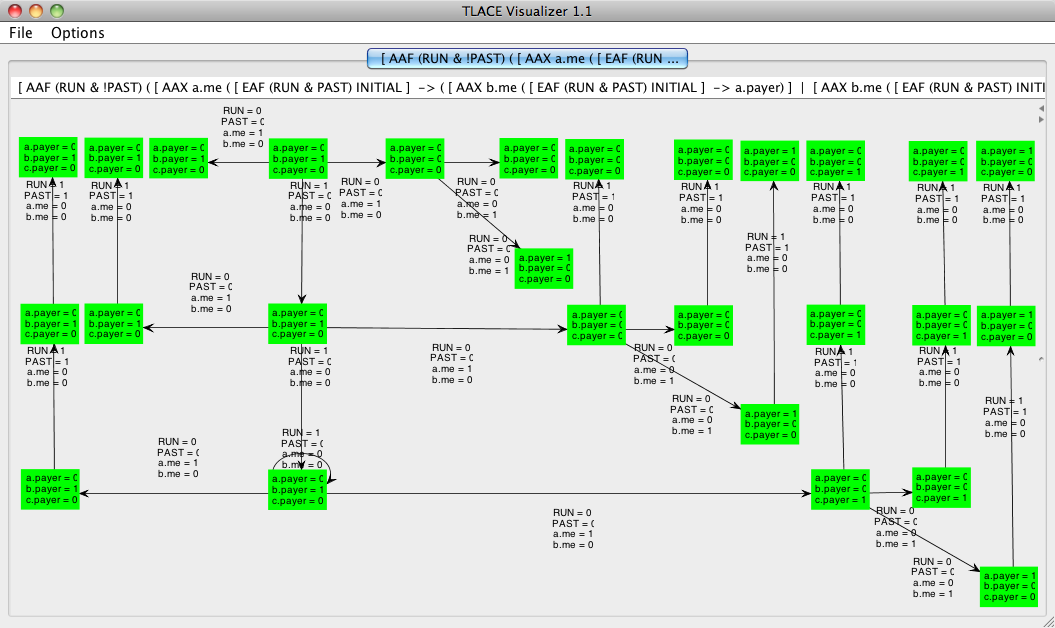}
			\caption{A counter-example for the property $\AF{\Kp{a}{\Kp{b}{a.payer}}}$ violated by the protocol of the dining cryptographers, as presented by the tool \toolname{}. The counter-example is expressed in terms of MTS and ARCTL. The inner values of states show who is the payer. The transitions labels show the transitions types: $RUN=1$ for temporal transitions, $PAST=1$ for reverse temporal transitions and $ag.me=1$ for the epistemic transitions of agent $ag$.}
			\label{figure:knowsknows}
		\end{figure}

		A screenshot of a counter-example for this property is given in Figure~\ref{figure:knowsknows}. The counter-example features many different branches, due to the nested operators and the disjunction resulting from $\Kp{ag}{\phi}$. This complexity increases when the number of nested epistemic operators increases: the counter-example for the property $\AF{\Kp{b}{\Kp{a}{\Kp{b}{a.payer}}}}$ contains $75$ TLACE nodes and $37$ branches; the counter-example for the same property with $8$ nested $\Kp$ operators would contain $195$ TLACE nodes and $97$ branches.
		
		While this increase still remains linear in terms of the length of the property, theory predicts (and experiments confirm) that the number of nodes of a tree-like counter-example for $\mathcal{M}, s \models \phi$ can be $O(|S|^{|\phi|})$ in the worst case, where $|S|$ is the number of states of the model $\mathcal{M}$ and $|\phi|$ the length of the violated property.  Indeed, for an ARCTL formula of depth $D$ (for example $\EaG{true}{\EaG{true}{...~\EaG{true}{b}}}$), the top-level branch in the counter-example may have up to $O(|S|)$ nodes, each of which carrying a branch with a counter-example of depth $D-1$, giving a total of $O(|S|^D)$ nodes.
		Consequently, the time needed to generate a tree-like annotated counter-example is at least in $O(|S|^{|\phi|})$ since the tool has to generate all nodes of the counter-example.  The exact performance depends on the complexity of BDD-based re-construction of the execution tree of the counter-example, amortized through the use of memoizing, which is beyond the scope of this analysis.

		\subsection{Towards Interactive Witness Generation}
		\label{section:interactive}
		
		As the size of the model and the complexity of the property grow, the generation of a counter-example may become intractable. A proposed solution, still to be investigated, is to generate the counter-examples in a lazy manner: instead of generating all the information in one batch, the tool outputs an initial state or an initial prefix of the counter-example and the user can ask the system to extend the parts of the counter-example that are most relevant to his understanding of the reported situation. 
		This interactive, incremental approach can also handle witnesses of universal operators: the user will be offered to ask for the expansion of selected branches, rather than being provided with the expansion of all branches.

		In such an approach, the Tool plays a game with the User where the Tool tries to show that the property is violated while the User tries to show that it is satisfied. The Tool will be responsible of showing witnesses for existential operators---by giving adequate branches---while the User will attempt to refute witnesses for universal operators (and fail, if the universal property indeed holds). 
		This approach will require a two-way interaction between the visualizer and the model checker (NuSMV), through which the visualizer will drive the incremental witness generation in the model checker.  The model checker will obviously have to be extended to support those incremental capabilities.
		This approach is related to game-based model checking, as developed e.g. in ~\cite{Stevens-Stirling-98,Alur-Henzinger-others-98}.

	\section{Related Work}
	\label{section:related-work}
			
		Other authors propose structures similar to tree-like annotated counter-examples to provide useful information about a violation.
		
		For example, Gurfinkel and Chechik generate proof-like counter-examples for CTL properties violated by Kripke structures~\cite{Gurfinkel-Chechik-03}. These counter-examples are based on a proof of the violation and are composed of states to which parts of the proof are linked. The proof steps are mechanically derived from the structure of the property and the counter-example, so a similar result could be produced on TLACEs by the visualization tool. Note that this would invert the process, since Gurfinkel and Chechick generate the counter-example from the proof and not the other way round.
		
		Shoham and Grumberg propose a game-based framework for CTL counter-example generation~\cite{Shoham-Grumberg-03}. Counter-examples are sub-graphs of the game-graph used to perform model checking. Each node of this graph is composed of a state of the model and a sub-formula of the property that it violates. This approach, similar to TLACEs and the proof-like counter-examples of~\cite{Gurfinkel-Chechik-03}, is applied to the context of incremental abstraction-refinement.
		The structure of these counter-examples is similar to TLACEs. Nevertheless, due to the granularity of the explanation, the number of steps to illustrate the violation---and then the number of nodes of the counter-example---is larger than for a TLACE. Such a counter-example for a given property $\phi$ is then larger than the corresponding TLACE for $\phi$, while giving the same information.
		
		Dong et al.\ define a framework to explore rich witness structures for modal $\mu$-calculus called \emph{evidences}~\cite{Dong-Ramakrishnan-others-03}. An evidence is a graph with nodes composed of a state of the system and a sub-formula of the property. As these evidences can be large, they develop a relational graph algebra to manipulate them, and provide an implementation. Evidences have a structure similar to counter-examples of Shoham and Grumberg~\cite{Shoham-Grumberg-03}. This framework could be adapted to explore multi-modal logics counter-examples like TLACEs.
		
		Meolic et al.\ propose another model for richer counter-examples~\cite{Meolic-Fantechi-others-04}. These counter-examples are automata accepting all finite linear traces of a given LTS violating a given ACTL (Action-based Computation Tree Logic) property. The tackled problem is not the same as the one of this paper since Meolic et al.\ only consider linear traces. Furthermore, they do not annotate their counter-examples.
		
		Some authors propose complementary approaches to analyze counter-examples and to extract their useful information.
		For example, Jin et al.\ annotate a linear counter-example into \emph{fated} and \emph{free-will} segments, representing the parts of the path where the environment can force the system to go to the error (fated segments) or where the system performs mistaken behavior and could avoid it (free-will segments)~\cite{Jin-Ravi-others-04}.
		This approach is complementary to any counter-example generation and could be applied to tree-like annotated counter-examples.
		Other authors like Groce and Visser~\cite{Groce-Visser-03} and Copty et al.~\cite{Copty-Irron-others-03} generate and check variations of a found (linear) counter-example to identify the critical parts that cause the violation. This approach is also complementary and could in principle be applied to tree-like counter-examples.
		
		Some people have applied SAT solving to verify CTL properties. For example, Penczek et al.\ describe an algorithm to transform an ACTL (the universal fragment of CTL) model checking problem into a SAT problem~\cite{penczek-wozna-others-02a}. The idea is to create a set of paths of length $k$ in the model connected together to form a branching counter-example. A solution to the SAT problem represents a viable counter-example. Nevertheless, they do not explicitly describe how to provide the counter-example to the user.
		
		In the domain of epistemic logics, MCK~\cite{Gammie-Meyden-04} and MCMAS~\cite{Lomuscio-Raimondi-06} are two tools that perform CTLK model checking. The first one, MCK, provides some debugging features like the export of the full graph of the system or the counter-example resulting from bounded model checking of a property. MCMAS gives also some debugging features: it presents a branching counter-example for a violated CTLK property, similar to TLACEs, but these counter-examples are not annotated. It also displays states information that can be filtered in terms of agents but does not give browsing features like the ones provided by \toolname{}.

	\section{Conclusion and Perspectives}
	\label{section:conclusion}
		
		This paper presents a structure, an algorithm and an implementation to represent, generate and explore \emph{tree-like annotated counter-examples} (TLACEs) for Action-Restricted CTL. These counter-examples are branching, explaining why an ARCTL property is violated by a model, but also recursively explaining why sub-formulas of the negation of the property are satisfied by the model. Furthermore, elements of these counter-examples are annotated to help their understanding. While these counter-examples explain ARCTL formulas violations, they become really useful to explain violations of richer branching logics like CTLK, that can be reduced to ARCTL.
		
		The algorithm uses sub-algorithms to generate paths in the model satisfying particular temporal operators and works recursively to explain why sub-formulas are satisfied by states of these paths.
		The implementation combines an extension of the NuSMV model checker 
		for generating TLACEs and a graphical tool for  displaying and inspecting them interactively.
		
		These counter-examples give more information about the violation than linear counter-examples and give a better understanding about the system. The annotations help the user to understand the structure of the counter-example.
		By nature, such branching counter-examples can become very large and their generation is computationally more costly than generating linear counter-examples.  The provided visualization tool is essential for conveniently and productively inspecting such large structures.	
				
		To be able to handle larger and more complex counter-examples, we are investigating an interactive incremental approach,
		where the tool provides an initial state of the model violating the property and the user can ask to expand branches of interest.
		This approach would also make it possible to provide useful witnesses for universal operators, by allowing the user to choose and simulate only selected branches.

	\bibliographystyle{eptcs}
	\bibliography{main.bib}
\end{document}